\documentclass[aps,twocolumn,showpacs]{revtex4}
\usepackage[english]{babel}
\usepackage{graphicx}
\usepackage{dcolumn}

\def\beq#1{\begin{equation}\label{#1}}
\def\eeq{\end{equation}}

\begin{document}
\title{High-order harmonic generation and Fano resonances}
\author{ V.V. Strelkov$^1$\footnote {e-mail:strelkov.v@gmail.com},
 M.A.Khokhlova$^{1,2}$, N.Yu Shubin$^3$}
\affiliation{$^1$A. M. Prokhorov General Physics Institute of the RAS, Moscow,  Russia \\
$^2$Faculty of Physics, M.V. Lomonosov Moscow State University, Moscow, Russia \\
$^3$Scientific Research Institute for System Studies of RAS, Moscow, Russia}

\begin{abstract}\noindent 
We present a high harmonic generation theory which generalizes the strong-field approximation to the resonant case, when the harmonic frequency is close to that of the transition from the ground to an autoionizing state of the generating system. We show that the line shape of the resonant harmonic is a product of the Fano-like factor and the harmonic line which would be emitted in the absence of the resonance.  The theory predicts rapid variation of the harmonic phase in the vicinity of the resonance. The calculated resonant harmonic phase is in reasonable agreement with recent measurements. Predicting the phase-locking of a group of resonantly-enhanced harmonics, our theory allows to study the perspectives of producing attosecond pulse train using such harmonics. 
\end{abstract}
\maketitle
\noindent

Although high-order harmonic generation (HHG) via interaction of intense laser pulses with matter provides a unique source of coherent femtosecond and attosecond pulses in the extreme ultraviolet (XUV),  the low efficiency of the process is a serious limit for its wide application. Using the resonances of the generating medium is a natural way to boost the efficiency, as was suggested already in the early HHG experimental~\cite{Toma_res} and theoretical~\cite{Roso,Platonenko} studies. Generation of high harmonics with frequencies close to that of the transition from the ground to an autoionizing state (AIS) of the generating particle were experimentally investigated in plasma media (for a recent review see~\cite{Ganeev,Ganeev_book}), and in noble gases~\cite{He,Xe-Kr}.

A number of theories describing HHG enhancement due to bound-bound
transitions were suggested~\cite{Becker_res,Maquet_res,Gaarde,Noble,Ishikawa}, and recently theories based on the specific properties of AIS were developed~\cite{Milosevic_res,IIvanov, Strelkov, Frolov}. These theories involve rescattering model~\cite{simple-man1,simple-man2} in which the HHG is described as a result of tunneling ionization, classical free electronic motion in the laser field, and recombination accompanied by the XUV emission upon the electron's return to the parent ion. In particular, in~\cite{Strelkov} one of us suggested a four-step resonant HHG model. The first two steps are the same as in the three-step model, but instead of the last step (radiative recombination from the continuum to the ground state) the free electron is trapped by the parent ion, so that the system (parent ion + electron) lands in the AIS, and then it relaxes to the ground state emitting XUV. 

Besides, there are several theoretical studies in which the HHG efficiency was calculated using the recombination cross-section. It was done heuristically~\cite{Plat-scatt} and analytically~\cite{Frolov-scatt} for the Coulomb interaction and generalizing the numerical results for the molecules~\cite{Lin}.

In this paper we suggest the high-order harmonic generation theory considering an AIS in addition to the ground state and free continuum treated in the theory for the non-resonant case~\cite{Lew}.
 We show that such accurate consideration verifies the model~\cite{Strelkov}. Moreover, we show that the intensity of the resonant HHG is described with a Fano-like factor that includes the scattering cross-section. However, in contrast to the previously suggested theories our approach allows also calculating the resonant harmonic's phase.   



We start with the time-dependent Schr\"odinger equation for an atom or ion in an external laser field linearly polarized along the $x$-axis:

\begin{equation}
i\frac{\partial}{\partial t}\Psi(\textbf{r},t) = \hat H  \Psi(\textbf{r},t), 
\label{SE}
\end{equation} 
where $\hat H = \left(-\frac{1}{2}\nabla^{2}+V(\textbf{r})-E(t)x\right) $ 
is the total Hamiltonian, and $\textbf{r}$ is the set of coordinate vectors of the electrons in the atom (ion). The wave function can be written as a sum of the ground state, unperturbed continuum and AIS:
\begin{equation}
\Psi= \Psi_{ground}+\Psi_{free}+c(t)\Psi_{AI}
\label{wave_func}
\end{equation}

To solve the Schr\"odinger equation~(\ref{SE}) we use the perturbation method. The wave function obtained in the absence of the AIS in the strong-field approximation (SFA)~\cite{Keldysh,Lew} is taken as the unperturbed solution: $\Psi_{0}= \Psi_{ground}+\Psi_{free}$. The solution $\Psi_{0}$ was found in~\cite{Lew} within the single-electron approximation neglecting the interaction of the free electron with the nucleus and other electrons:

\begin{equation}
\Psi_{0}(\textbf{r},t)= e^{iI_{p}t}\left(a(t)\varphi_{gr} (\textbf{r}) + \int d^{3}\textbf{v}b(\textbf{v},t)\chi (\textbf{r})\right),
\label{wave_func_Lew}
\end{equation}
where $\varphi_{gr}(\textbf{r})$ is the ground state and $\chi(\textbf{r})$ is a flat wave. This solution can be generalized to the multi-electron case as follows. Let $\textbf{r}_1$ is the radius-vector of the ''active'' electron and $\tilde \textbf{r}$ are radius-vectors of the other electrons: $\textbf{r}=\{\tilde \textbf{r}, \textbf{r}_1 \}$. Let $\chi(\textbf{r})=\chi_1(\textbf{r}_1) \tilde \varphi_{gr}(\tilde \textbf{r}) \exp(i \tilde I_p t)$ where $\tilde I_p$ is the ionization potential of the {\it parent ion}. Also, let us write formally $ V'(\textbf{r})=V(\textbf{r})- \tilde V (\tilde \textbf{r} )$,
where $\tilde V (\tilde \textbf{r} )$ is the part of the potential which depends only on $\tilde \textbf{r}$. Now the solution $\Psi_{0}(\textbf{r},t)$ can be found via the procedure similar to that used in~\cite{Lew}. Namely, neglecting the term $V'(\textbf{r}) \chi(\textbf{r})$ (but not the term $\tilde V (\tilde \textbf{r} ) \chi(\textbf{r})$ ) in the Schr\"odinger equation we find the equation describing $b(\textbf{v},t)$ coinciding with Eq.~(4) from~\cite{Lew}. Thus the obtained solution satisfies the equation:
\begin{equation}
i\frac{\partial}{\partial t}\Psi_{0} =
\hat H  \Psi_{0}-V'(\textbf{r})\Psi_{free}. 
\label{SE_Lew}
\end{equation} 

Neglecting the modification of the AIS by the laser field, the AIS can be written as a solution of the Schr\"odinger equation:

\begin{equation}
i\frac{\partial}{\partial t}\Psi_{AI} =\hat H(\textbf{r})\Psi_{AI}  
\label{SE_qs}
\end{equation} 

Since we are planning to calculate the dipole moment of the system which is naturally localized near the origin, below we neglect the outgoing part of this solution, and focus on its part localized near the origin. This part can be written using the {\it complex} energy~\cite{Gamov_funct,Manson,Mercouris,decay}:

\begin{equation}
\Psi_{AI}(\textbf{r},t)= \varphi(\textbf{r})\exp(-iWt). 
\label{SE_qs1}
\end{equation}

Here $\varphi(\textbf{r})$ is a stationary solution (in absence of the configuration interaction, see~\cite{Fano}), and the energy is $W=W_0-i \Gamma/2$, where $W_0$ is the real energy of the AIS and $\Gamma$ is the AIS  width (see~\cite{Fano}), 

\begin{equation}
\Gamma=2 \pi |V_{1}(v_r)|^2,
\label{Gamma}
\end{equation} 
where $V_{1}(\textbf{v})=<\chi(\textbf{v})   |V'(\textbf{r})|   \varphi >$ and $v_r=\sqrt{2 W_0}$. Note that $\Gamma=1/\tau$, $\tau$ is the lifetime of the AIS.

Using equations 
(\ref{SE})-(\ref{SE_qs}) we obtain:

\begin{equation}
i \dot c \Psi_{AI} = V'(\textbf{r}) \Psi_{free}
\label{SE_c}
\end{equation}

Multiplying this equation by $\varphi^* (\textbf{r})$ and integrating over $\textbf{r}$ we have:

\begin{equation}
\dot{c}(t)=-ie^{i(W+I_{p})t}\int d^{3}\textbf{v} b(\textbf{v},t)V^*_{1}(\textbf{v}).
\label{c_t}
\end{equation}

The solution of this equation is
\begin{equation}
c(t)=-i\int_{-\infty}^{t}dt'e^{i(W+I_{p})t'}\int d^{3}\textbf{v}b(\textbf{v},t')V^*_{1}(\textbf{v}).
\label{c}
\end{equation}

We can see that the matrix element $V_{1}(\textbf{v})$ is the parameter which determines the amplitude of the AIS. Thus, this state is a small perturbation of the continuum when this parameter is small (in atomic units). Note that a similar requirement appears in the Fano theory~\cite{Fano} and its applications as the requirement for the resonance to be well-isolated in the continuum: $\Gamma<<W_0$. This condition is usually valid for the autoionizing states of atoms and ions. 

Below we find the time-dependent dipole moment of the system $\mu(t)=<\Psi(t)|x|\Psi(t)>$. Substituting the wave function as (\ref{wave_func}), neglecting the contribution of the continuum-continuum transitions to the dipole moment (as in~\cite{Lew}), and contribution of the continuum-AIS transitions to the dipole moment (both assumptions are valid in case of low population of the continuum) we obtain

\begin{equation}
\mu(t)=<\Psi_{ground}|x|\Psi_{free}>+<\Psi_{ground}|x|c(t)\Psi_{AI}>+ c.c.
\label{dip}
\end{equation}

The first term describes HHG in the absence of the resonances of the generating system (see equation (6) in~\cite{Lew}):
\begin{equation}
\mu_{nr}(t)=\int d^{3}\textbf{v}b(\textbf{v},t)d^{*}_{nr}(\textbf{v}) +c.c.
\label{dip_nr}
\end{equation}
where $d_{nr}(\textbf{v})=<\chi(\textbf{v})|x|\varphi_{gr}>$ is the dipole matrix element of the continuum-ground state transitions. The second term in Eq.~(\ref{dip}) describes the effect of the resonance on the harmonic generation:
$$\mu_{r}(t)=e^{-i(W+I_{p})t}c(t)d^{*}_{r} + c.c.$$ 
where $d_{r}=<\varphi|x|\varphi_{gr}>$ is the dipole matrix element of the AIS-ground state transition. Substituting here Eq.~(\ref{c}) we obtain
\begin{equation}
\mu_{r}(t)=-id^{*}_{r}\int_{-\infty}^{t}dt'e^{i(W+I_{p})(t'-t)}\int d^{3}\textbf{v}b(\textbf{v},t')V^*_{1}(\textbf{v})
 + c.c.
\label{dip_r}
\end{equation}
where $b(\textbf{v},t')$ (the wave function amplitude of the electron in the free continuum) is the same as that in the Lewenstein's theory (see equation (5) in~\cite{Lew}).

We transform the equation (\ref{dip_r}), so that it is written using non-resonant contribution~(\ref{dip_nr}). We suppose that $V^*_{1} (\textbf{v})$ and $d^{*}_{nr}(\textbf{v})$ are smooth functions of the velocity, so they can be taken outside the integral at $v=v_{r}$. Introducing $\tau'=t-t'$ we have
\begin{equation}
\mu_{r}(t)=-i\frac{V^*_{1}(v_{r})d^{*}_{r}}{d^{*}_{nr}(v_{r})}
\int^{\infty}_{0}d\tau' e^{-i(W+I_{p})\tau'}\mu_{nr}(t-\tau')
\label{dip_r_}
\end{equation}
Thus, the spectrum [$f(\omega)=\int_{-\infty}^{\infty} f(t) \exp(i\omega t)dt$] of the resonant contribution to the dipole moment is
\begin{equation}
\mu_{r}(\omega)=\frac{V^*_{1}(v_{r})d^{*}_{r}}{d^{*}_{nr}(v_{r})}\frac{\mu_{nr}(\omega)}{\omega-(W_{0}+I_{p})+i\frac{\Gamma}{2}},
\label{dip_w}
\end{equation}
where $\mu_{nr}(\omega)$ is the spectrum of the non-resonant contribution given by Eq. (8) in~\cite{Lew}. Introducing the detuning from the resonance $\Delta \omega=\omega-(W_0+I_p)$, we obtain the spectrum of the total dipole moment of the system~(\ref{dip}) taking into account~(\ref{dip_w}):
\begin{equation}
\begin{array}{l}
\mu(\omega)=\mu_{nr}(\omega) F(\omega),
\\
F( \omega)=\left[1+Q \frac{\Gamma/2}{\Delta \omega+i \Gamma/ 2}\right],
\label{dip_w_fin}
\end{array}
\end{equation}
where

\begin{equation}
Q=\frac
{V^*_{1}(v_{r})d^{*}_{r}}
{d^{*}_{nr}(v_{r})\Gamma/2}
\label{Q}
\end{equation}


The complex conjugate parameter $Q^{*}$ is close to the Fano parameter $q$ (which can be real or complex, see~\cite{complex_Fano_1,complex_Fano_2,complex_Fano_3}):

\begin{equation}
Q^{*} \approx q=\frac{<\Phi|x|i>}{\pi V^{*}_{E}<\psi_E|x|i>}
\end{equation} 

The approximate character of the equality is due to the following facts:

- the state $\Phi$ is different from $\varphi$, see Eq. (17) in~\cite{Fano};

- the state $\psi_E$ is different from the free wave $\chi$, see ~\cite{Fano} for more details.

The factor $|F(\omega)|^2$ describes the line profile, which coincides with the Fano profile when $Q=q-i$. Note that the phase $2\arg(q-i)$ is introduced in~\cite{Pfeifer} as a characteristic of the process dynamics described by the Fano line. The condition $\Gamma<<1$, ensuring the applicability of our approach, in the case of real $Q$ corresponds to $Q>>1$, therefore, the difference in line shape $|F(\omega)|^2$ from the Fano profile is small in this case.


In order to apply Eq.~(\ref{dip_w_fin}) for certain resonances, the complex values of $d_r$, $d_{nr}$, $V_1$ should be known.

We have:
$|d_r|^2=\frac{f_{osc}}{2 (I_p+W_0)}$
where $f_{osc}$ is the oscillator strength of the transition. Note that $f_{osc}$ also defines  the resonant photoionization cross-section, so our findings agree with those in the published studies describing the HHG intensity via photoionization cross-section~\cite{Plat-scatt, Frolov-scatt,Lin,Smirnova}.  

The matrix element $d_{nr}$ was found in~\cite{Lew} for different binding potentials. Note that the normalization of the free wave in~\cite{Lew} and~\cite{Fano} is different. If the normalization of~\cite{Fano} is used (providing Eq.~(\ref{Gamma})), the matrix elements $d_{nr}$ found in ~\cite{Lew} should be multiplied by $1/\sqrt{v_r}$.

Finally $|V_1|^2$ can be found from $\Gamma$ (see Eq.~(\ref{Gamma})) which was calculated or measured for many transitions. However, calculation of the phase of $V_1$ is a separate problem which is discussed in the Appendix, see supplementary materials.  

\begin{figure}  
\centering
\includegraphics [width=0.7\columnwidth] {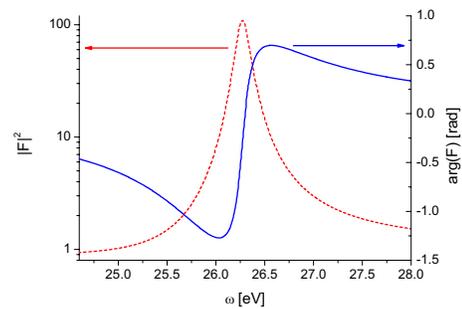}
\caption{(color online) Absolute value and argument of the factor $F(\omega)$ in Eq.~(\ref{dip_w_fin}), calculated for the \hbox{4d$^{10}$5s$^2$5p $^2$P$_{3/2} \leftrightarrow$  4d$^{9}$5s$^2$5p$^2$~($^1$D)$^2$D$_{5/2}$} transition in Sn$^+$, the transition frequency is 26.27 eV thus close to the 17-th harmonic of Ti:Sapp laser.}
\label{fig1}
\end{figure}


In Fig.~\ref{fig1} we present the absolute value and the phase of factor $F(\omega)$, calculated for the transition in Sn$^+$, which is important because of the comparison with experiments discussed below. In the figure one can see the line asymmetry and a rapid phase variation in the vicinity of the resonance. From Eq.~(\ref{dip_w_fin}) we can see that the slope of the phase at the resonance for high $|Q|$ is approximately the doubled lifetime of the AIS: $\partial [\arg(F)]/ \partial \omega \approx 2/\Gamma=2\tau$. This slope corresponds to the delay in the emission of the resonant harmonic. The presence of this delay confirms the four-step mechanism of the resonant HHG~\cite{Strelkov}, as it was first pointed out in the numerical studies~\cite{Tudorovskaya}.

\begin{figure}  
\centering
\includegraphics [width=0.8\columnwidth] {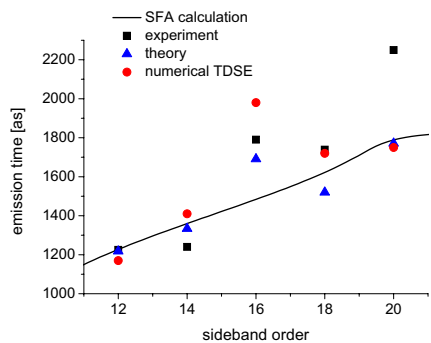}
\caption{(color online) Calculated and measured emission time for high-order harmonics generated in tin plasma plumes.  Triangles show calculated results based on the present theory, the other results are from Ref.~\cite{Haessler-atto}: dots and squares show the numerical results and those of the \textsc{rabbit}  measurements, respectively, and the black line shows the results of a SFA calculation.}
\label{fig2}
\end{figure}

Recently the first temporal characterization of the attosecond emission from tin plasma in resonant conditions was performed~\cite{Haessler-atto} using a RABBIT technique~\cite{Paul2001Observation}. It was shown that the resonance considerably changes the relative phase of the neighboring harmonics. The emission time $\tau_e$ found with RABBIT for sideband $q$ is linked to the
phases $\varphi_{q-1}$ and $\varphi_{q+1}$ of two neighboring harmonic as $\tau_e=(\varphi_{q+1}-\varphi_{q-1})/2\omega_\mathrm{0}$, where $\omega_\mathrm{0}$ is the laser frequency. Thus the change of the emission time due to the resonance is (see Eq.~(\ref{dip_w_fin})): $\Delta\tau_e=\{\arg[F((q+1)\omega_0)]-\arg[F((q-1)\omega_0)]\}/2\omega_\mathrm{0}$. The emission time calculated with this correction is shown in Fig.~\ref{fig2} together with the experimental and numerical results from~\cite{Haessler-atto}. We can see that for the sidebands (SB) far from the resonance (SB12 and SB14) both the theory and the experiment show emission time in agreement with the SFA prediction, whereas near the resonance (SB16 and SB18) this is not the case. For SB16 our theory agrees with the experiment. The change of the $\tau_e$ from SB 16 to SB18 is negative both experimentally and theoretically, but the measured change is smaller. Note that experimentally the RABBIT signal for the SB18 was not very stable~\cite{Haessler-atto}.  

\begin{figure}  
\centering
\includegraphics [width=0.9\columnwidth] {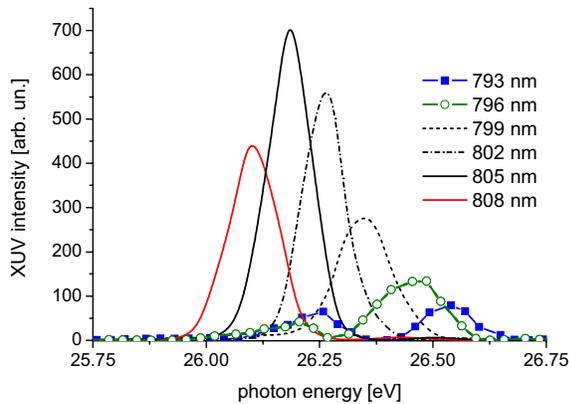}
\caption{(color online) The calculated harmonic spectrum in the vicinity of the resonance for different fundamental wavelengths, leading to different detunings from the resonance. The resonant transition is the same as in Fig.~\ref{fig1}.}
\label{fig3}
\end{figure}


The harmonic phase variation due to the factor $F(\omega)$ in the vicinity of a wide (i.e. covering several harmonics) resonance can be used for compensation of the attochirp (variation of $\tau_e$ as a function of the harmonic order). Namely for the resonant harmonics {\it above} the resonance the attochirp is compensated for the short electronic trajectory, and for those {\it below} the resonance it is compensated for the long one. The variation of the additional emission time is $\partial \Delta \tau_e / \partial \omega = \partial^2 \arg[F(\omega)]/\partial \omega^2 \sim 1/\Gamma^2$. This estimate shows that, in particular, resonant HHG in Xe using 1-2 $\mu m$ pump is a good candidate for attochirp compensation via the resonance in Xe at approximately 100 eV.

In Fig.~\ref{fig3} we present the spectrum of the resonant 17-th harmonic calculated using numerical TDSE solution as described in~\cite{Haessler-atto,Ganeev_Sn} averaged for the laser intensities up to $0.8 \times 10^{14}$ W/cm$^2$, the laser pulse duration is 50 fs. One can see that different detunings from the resonance lead to different peak harmonic intensities and, what is more interesting, to different harmonic line shapes: for the 793nm, 796nm, and 808nm fundamental the harmonic line consists of two peaks; it is known for the non-resonant HHG that these peaks can be attributed to the contributions of the short and the long electronic trajectories, see~\cite{Zair} and references therein. In the figure we can see that the long trajectory's contribution is in general weak, but it becomes more pronounced when its frequency is closer to the exact resonance, as it is the case for the 793nm fundamental. These results illustrate the fact that the harmonic line shape can be well-understood via the factorization of the harmonic signal described by Eq.~(\ref{dip_w_fin}).  This straightforward factorization is a remarkable fact, considering the complexity of the dynamics of {\it both} free electronic wave-packet and the AIS, which determine the harmonic line shape.



Above we have considered a single AIS. However, our perturbation theory can be easily generalized for the case of multiple non-overlapping autoionizing  (AI) states, keeping the assumption that the influence of these states on the free electronic wave-packet $\Psi_{free}$ remains small. Namely, to take into account several AI states, the terms corresponding to each state with its specific $Q$, $\Delta\omega$ and $\Gamma$ should be added in the brackets in the right part of Eq.~(\ref{dip_w_fin}).


As we have already mentioned, we neglect the influence of the laser field on the AIS. However, in certain cases this influence can be important as it is shown in~\cite{Pfeifer,AI_in_field_1,AI_in_field_2,AI_in_field_3}. Accurate consideration of the laser field requires essential modification of the theory. However, one can suppose that the channels of the auto- and photo-ionization do not interfere because they have different final states: that of the autoionization is the free state with energy $W_0$, and that of the photoionization is the free state with energy $W_0+n \omega$ appearing upon absorption of $n$ photons. Note that the assumption of the absence of interference neglects such process as the ''field-assisted tunneling\grqq~\cite{assisted_1,assisted_2} with $n=0$; however, in the case of multiphoton ionization the contribution of the process with $n=0$ should be small. Within this assumption the width of the state in the laser field is $\Gamma'=\Gamma+w_{ph}$, where $w_{ph}$ is the photoionization rate. To estimate this rate we suppose that the photoionization occurs in the absence of the configuration interaction. In this case for the doubly excited state of the atom (for instance, in He) the ''ionization energy''~$I_{AI}$ is the difference of the energy of the system ''excited ion + free electron'' and the doubly excited state of the atom (for 2s2p state of He this is 65.17eV-59.91eV=5.26eV, see the energy structure of He, for instance, Fig.1 in~\cite{Starace}). Knowing $I_{AI}$ one can calculate the photoionization rate using the Keldysh formula~\cite{Keldysh}. Moreover, the time-dependence of $w_{ph}$ due to the temporal variation of the laser field can be taken into account. In this case the time-dependence of $\Gamma$ does not allow analytical integration in Eq.~(\ref{dip_r}), and thus the dipole moment cannot be presented in the factorized form~(\ref{dip_w_fin}). 


In conclusion, in this paper we present the theory which generalizes the SFA approach for HHG to the resonant case, considering an AIS in addition to the ground state and the free continuum state; the latter two states are treated in the same way as in the theories developed for the non-resonant case. The main result is given by Eq.~(\ref{dip_w_fin}) presenting the resonant harmonic line as a product of the Fano-like factor and a harmonic line which would be emitted in the absence of the AIS. Our theory allows calculating not only the resonant harmonic intensity, but also its phase. We show that there is a rapid variation of the phase in the vicinity of the resonance. Our calculations reasonably agree with the RABBIT harmonic phase measurements. Our theory predicts that in the case of a resonance covering a group of harmonics the resonance-induced phase variation can compensate for the attochirp in a certain spectral region. The natural perspective of our studies is taking into account the influence of the laser field on the AIS.

The authors acknowledge fruitful discussions with D. Kilbane, M. Fedorov, A. Magunov and P. Sali\'eres. This study was supported by RFBR (grant N 12-02-00627-a), Ministry of Education and Science of RF (grant N MD-6596.2012.2), and the Dynasty Foundation.

\end{document}